# A Transform Method of a Force Curve Obtained by Surface Force Apparatus to the Density Distribution of a Liquid on a Surface: An Improved Version

**Ken-ichi Amano[a] and Eisuke Tanaka[b]**

[a]*Department of Energy and Hydrocarbon Chemistry, Graduate School of Engineering, Kyoto University, Kyoto 615-8510, Japan.*

[b]*Undergraduate School of Industrial Chemistry, Faculty of Engineering ,Kyoto University, Kyoto 615-8510, Japan.*

Author to whom correspondence should be addressed: Ken-ichi Amano.

Electric mail: amano.kenichi.8s@kyoto-u.ac.jp

**Abstract**

We propose a transform method from a force curve obtained by a surface force apparatus (SFA) to a density distribution of a liquid on a surface of the SFA probe. (We emphasize that the transform method is *a theory for the experiment*.) In the method, two-body potential between the SFA probe and the solvent sphere is modeled as the soft attractive potential with rigid wall. The model potential is more realistic compared with the rigid potential applied in our earlier work. The introduction of the model potential is the improved point of the present transform method. The transform method is derived based on the statistical mechanics of a simple liquid where the simple liquid is an ensemble of small spheres. To derive the transform method, Kirkwood superposition approximation is used. It is found that the transformation can be done by a sequential computation. It is considered that the solvation structure can be obtained more precisely by using the improved transform method.

## 1. Introduction

In the short letter, we propose a method for transforming a force curve obtained by surface force apparatus (SFA) [1,2] to a (number) density distribution of a liquid on a SFA probe. We call the density distribution solvation structure. In our earlier work [3], the transformation is performed by modeling the two-body potential between the SFA probe and the solvent sphere as the rigid potential. However, the approximation of the rigid potential is very rough. It should be improved to a more realistic model. Hence, in the present study, the two-body potential is modeled as the soft attractive potential with rigid wall. The model potential is considered to be more realistic except for the rigid wall.

The transform method is derived based on the statistical mechanics of simple liquids. In the derivation, the solvent is modeled as an ensemble of small spheres, and two-body potential of which is arbitrary. Two-body potential between the SFA probe and the solvent sphere is modeled as soft attractive potential with rigid wall. The rigid wall is applied in order to clearly determine the contact point between them. When the transform method is used, the force curve is divided into two forces beforehand. One is the solvation force and the other is two-body force between the SFA probe and the sample surface. (We simply call both the SFA probe and the solid sample the solid plates. The SFA probe and the sample are set to be the same in the derivation.) The solvation force is used as an input datum for calculation of the solvation structure. Since the repulsive area is approximated to the rigid wall, the maximum position of the first layer of the solvation force between the SFA probe and the sample surface is interpreted as that the sandwiched solvent spheres are just contacting both the SFA probe and sample surfaces. The introduction of the rigid wall in the repulsive area make the practical use of the transform method accessible because of the simplicity of the determination of the contact point between surfaces of the solid plate and solvent sphere.

To derive the transform method, Kirkwood superposition approximation [4-6] is used. It is found that the solvation structure can be obtained by a sequential computation. Firstly, the number density on the contact point between the solid plate and solvent sphere is calculated. Secondly, the number density on a position that slightly far from the contact point is calculated by using the previous output, the number density on the contact point. Thirdly, the number density on a position that slightly far from the second position is calculated by using the previous outputs. The number densities of the other positions are calculated by the sequential computation written above. It is considered that the solvation structure can be obtained more precisely by using the present transform method. In Chapter 2, the derivation process of the improved transform method is demonstrated, and in Chapter 3, short conclusions are presented.

## 2. Theory

In this chapter, derivation process of the transform method is demonstrated. To start the demonstration, we introduce following conditions:

(I) The solvent considered here is a simple liquid, an ensemble of the small spheres. The diameter of the solvent sphere is $d_S$. Two-body potential between the solvent spheres is arbitrary.
(II) Two-body potential between the SFA probe and the solvent sphere has an attractive soft potential with repulsive rigid wall.
(III) Cylindrical solids 1 and 2 are immersed in the simple liquid. (Cylindrical solids 1 and 2 represent the solid sample and the SFA probe, respectively.) The cylindrical solids have the same shapes. (The cylindrical solids can be alternated with cuboidal solids in the theory. Furthermore, the shapes are not required to be the same in theory. However, here we use only these cylindrical solids to simplify the demonstration of the derivation process.)
(IV) The circular surfaces of the cylindrical solids are facing each other, and the circular surfaces are vertical to the $z$-axis.
(V) The origin of the whole system is set at the center of the facing surface of cylindrical solid 1. Position of the cylindrical solid 1 is fixed, whereas cylindrical solid 2 changes its position along the $z$-axis.
(VI) Surface areas of the circular surfaces are sufficiently large compared to diameter of the solvent sphere. This enables us to ignore the solvated spheres near the edges of the cylindrical solids.
(VII) The lateral surfaces of the cylindrical solids are horizontal to the $z$-axis, and hence the solvation force along the $z$-axis is never generated from the solvated spheres near the lateral surfaces.
(VIII) Heights of the cylindrical solids are sufficiently large, so that the solvation structures on the backward surfaces (the non-facing surfaces) are never destroyed in any separation between the cylindrical solids.
(IX) Any phase transitions occurring in a local space sandwiched between the facing surfaces (e.g., drying effect) are ignored in the theory.

From here, we give a description of the derivation process by considering the above conditions. In the SFA experiment, the force between the two surfaces is measured, and the solvation force can be picked out from the crude force by subtracting the two-body force between the SFA probe and the solid sample. The two-body force can be theoretically calculated from that of the two-body potential or measured by SFA experiment in the air (vacuum). The solvation force along the $z$-axis being $f_{\text{sol}}$ (solvation force acting on the cylindrical solid 2) has an exact relationship with the number density distribution of the solvated spheres, which is expressed as [7]

$$f_{\text{sol}}(s) = A \int_{-\infty}^{\infty} \rho(z;s) \frac{\partial u_2(z;s)}{\partial z} dz, \qquad (1)$$

where $A$ represents the facing surface area of the cylindrical solid 2. $\rho(z;s)$ is the number density of the solvent at $z$, and $s$ is the separation between the facing surfaces. $u_2$ is the two-body potential between cylindrical solid 2 and the solvent sphere. Eq. (1), derived by considering an infinitesimal change of *the separation between the cylindrical solids*, is an exact expression in a classical manner. It is derived by using statistical mechanics of a simple liquid. Although Eq. (1) of the solvation force seems to originate from the sum of the solvation energy, it originates from the sum of the solvation free energy. The solvation energy $f_{\text{sol}}$ corresponds to positional differential of the solvation free energy of a pair of the solid plates. That is, a factor of the solvation entropy is also surely considered (contained) in Eq. (1). By the way, Eq. (1) is strictly consistent with the contact theorem [8-10]. (The contact theorem explains the pressure on a wall, the derivation of which is performed by an infinitesimal change of *the volume of a system or a solute*.) To connect the solvation force and the solvation structures on cylindrical solids 1 and 2, we take advantage of the Kirkwood superposition approximation [4-6] and express $\rho$ as

$$\rho(z;s) \approx \rho_0 g_1(z) g_2(z-s). \qquad (2)$$

Here, $\rho_0$ is the bulk number density of the solvent (which is constant), and $g_i$ ($i = 1$ or 2) is a pair correlation function between the cylindrical solid $i$ and solvent. $g_i$ is the so-called a normalized number density of the solvent and/or a solvation structure around the cylindrical solid $i$. Applying this approximation, Eq. (1) is rewritten as

$$f_{\text{sol}}(s) = A\rho_0 \int_{-\infty}^{\infty} g_1(z) g_2(z-s) u_2'(z-s) dz. \qquad (3)$$

The origin of $g_1$ is placed at the center of the whole system, whereas the origin of $g_2$ is placed at the center of the facing surface of the cylindrical solid 2. The origin of $u_2$ is the same as that of $g_2$. $u_2$' represents the partial differentiation of $u_2$ with respect to $z$. Considering the conditions, Eq. (3) can be rewritten as

$$f_{\text{sol}}(s) = A\rho_0 \int_0^s g_1(z) g_2(z-s) u_2'(z-s) dz - PA, \qquad (4)$$

where $P$ represents the pressure on a wall in the solvent. Since two-body potential between the cylindrical solid 2 and the solvent sphere being $u_2$ has the attractive soft potential and the rigid wall, the value of $\exp[-u_2/(k_B T)]$ is expressed as

$$\exp[-u_2/(k_B T)] = 0 \quad \text{for overlapped points,} \tag{5a}$$
$$\exp[-u_2/(k_B T)] = b > 1 \quad \text{for contact point,} \tag{5b}$$
$$\exp[-u_2/(k_B T)] \geq 1 \quad \text{for the other points,} \tag{5c}$$

where $k_B$ and $T$ are the Boltzmann constant and absolute temperature, respectively. The partial differentiation of $u_2$ with respect to $z$ can be expressed as

$$\frac{\partial u_2(z-s)}{\partial z} = -k_B T \exp[u_2(z-s)/(k_B T)] \frac{\partial \exp[-u_2(z-s)/(k_B T)]}{\partial z}. \tag{6}$$

Hence, the partial differentiation in the very vicinity of the contact point (VCP) is expressed as

$$\left. \frac{\partial u_2(z-s)}{\partial z} \right|_{\text{VCP}} = k_B T \delta[z - (s - d_S/2)], \tag{7}$$

where $\delta$ is the delta function. By substituting Eq. (7) into Eq.(4), $f_{\text{sol}}(d_S)$ is calculated to be

$$f_{\text{sol}}(d_S) = A k_B T \rho_0 g_1(d_S/2) g_2(-d_S/2) - PA, \tag{8}$$

where the values of $g_1(d_S/2)$ and $g_2(-d_S/2)$ both are the normalized number density at the contact points. They can be represented as $g_C$ where the subscript C denotes the contact point. Therefore, $g_C$ can be obtained as follows:

$$g_C = \sqrt{\frac{f_{\text{sol}}(d_S) + PA}{A k_B T \rho_0}}. \tag{9}$$

Secondly, we consider $f_{\text{sol}}(d_S+\Delta s)$ in order to obtain $g_1(d_S/2+\Delta s)$, where $\Delta s$ is the sufficiently small separation. $f_{\text{sol}}(d_S+\Delta s)$ is expressed as

$$f_{\text{sol}}(d_S + \Delta s) = A\rho_0 g_1(d_S/2) g_2(-d_S/2 - \Delta s) u_2'(-d_S/2 - \Delta s) \Delta s$$
$$+ A k_B T \rho_0 g_1(d_S/2 + \Delta s) g_2(-d_S/2) - PA. \tag{10}$$

In the conditions written above, the cylindrical solids 1 and 2 are the same things. Thus, Eq. (10) is rewritten as

$$f_{\text{sol}}(d_S + \Delta s) = A\rho_0 g_C g_1(d_S/2 + \Delta s)u_2'(-d_S/2 - \Delta s)\Delta s$$
$$+ Ak_B T \rho_0 g_1(d_S/2 + \Delta s)g_C - PA. \tag{11}$$

Hence, $g_1(d_S/2+\Delta s)$ is calculated to be

$$g_1(d_S/2 + \Delta s) = \frac{f_{\text{sol}}(d_S + \Delta s) + PA}{A\rho_0 g_C[k_B T + u_2'(-d_S/2 - \Delta s)\Delta s]} \tag{12}$$

Thirdly, we consider $f_{\text{sol}}(d_S+n\Delta s)$ in order to obtain $g_1(d_S/2+n\Delta s)$, where $n$ represents arbitrary natural number. $f_{\text{sol}}(d_S+n\Delta s)$ is expressed as

$$f_{\text{sol}}(d_S + n\Delta s) = A\rho_0 g_1(d_S/2)g_2(-d_S/2 - n\Delta s)u_2'(-d_S/2 - n\Delta s)\Delta s$$
$$+ A\rho_0 g_1(d_S/2 + \Delta s)g_2(-d_S/2 - (n-1)\Delta s)u_2'(-d_S/2 - (n-1)\Delta s)\Delta s$$
$$+ A\rho_0 g_1(d_S/2 + 2\Delta s)g_2(-d_S/2 - (n-2)\Delta s)u_2'(-d_S/2 - (n-2)\Delta s)\Delta s$$
$$+ A\rho_0 g_1(d_S/2 + 3\Delta s)g_2(-d_S/2 - (n-3)\Delta s)u_2'(-d_S/2 - (n-3)\Delta s)\Delta s$$
$$\cdot$$
$$\cdot$$
$$\cdot$$
$$+ A\rho_0 g_1(d_S/2 + (n-1)\Delta s)g_2(-d_S/2 - \Delta s)u_2'(-d_S/2 - \Delta s)\Delta s$$
$$+ Ak_B T \rho_0 g_1(d_S/2 + n\Delta s)g_2(-d_S/2) - PA. \tag{13}$$

The form of Eq. (13) can be simplified as follows:

$$f_{\text{sol}}(d_S + n\Delta s) = A\rho_0 g_C g_2(-d_S/2 - n\Delta s)u_2'(-d_S/2 - n\Delta s)\Delta s$$
$$+ A\rho_0 \sum_{i=1}^{n-1} g_1(d_S/2 + i\Delta s)g_2(-d_S/2 - (n-i)\Delta s)u_2'(-d_S/2 - (n-i)\Delta s)\Delta s$$
$$+ Ak_B T \rho_0 g_1(d_S/2 + n\Delta s)g_C - PA. \tag{14}$$

Therefore, $g_1(d_S/2+n\Delta s)$ is calculated to be

$$g_1(d_S/2 + n\Delta s) = \frac{f_{\text{sol}}(d_S + n\Delta s) + PA - A\rho_0 \sum_{i=1}^{n-1} W(i)}{A\rho_0 g_C[k_B T + u_2'(-d_S/2 - n\Delta s)\Delta s]}, \tag{15}$$

where

$$W(i) = g_1(d_S/2 + i\Delta s)g_2(-d_S/2 - (n-i)\Delta s)u_2'(-d_S/2 - (n-i)\Delta s)\Delta s. \tag{16}$$

The calculation process of the solvation structure is as follows: (i) Calculate $g_C$ from Eq. (9); (ii) Calculate $g_1(d_S/2+\Delta s)$ by using the previous output $g_C$ from Eq. (12); (iii) Calculate $g_1(d_S/2+2\Delta s)$ by using the previous outputs $g_C$ and $g_1(d_S/2+\Delta s)$ from Eq. (15); (iv) Calculate $g_1(d_S/2+3\Delta s)$ by using the previous outputs $g_C$, $g_1(d_S/2+\Delta s)$, and $g_1(d_S/2+2\Delta s)$ from Eq. (15); (v) Calculation of $g_1(d_S/2+n\Delta s)$ is performed by using the previous outputs from Eq. (15). Thus, the calculation of the solvation structure is performed by the sequential computation.

In this section, we argue about the width of $\Delta s$. Seeing Eq. (15), one can realize that its denominator must be an positive value, because $g_1(d_S/2+n\Delta s)$ and its numerator are positive values. That is, the following magnitude relation must be realized:

$$k_B T + u_2'(-d_S/2 - n\Delta s)\Delta s > 0. \qquad (17)$$

Hence, the value of $\Delta s$ must be within

$$0 < \Delta s < -k_B T/u_2'(-d_S/2 - n\Delta s). \qquad (18)$$

In most cases, slope of the attractive soft potential $u_2$ becomes steeper as the position approaches the rigid wall. Therefore, the numerical range is rewritten as (see Eq. (12))

$$0 < \Delta s < -k_B T/u_2'(-d_S/2 - \Delta s). \qquad (19)$$

When the sequential computation is performed, one must follow Eq. (18) or (19). Generally, the numerical range of Eq. (19) is more severe than that of Eq. (18). Thus, in that case, one must follow Eq. (19).

There is another point that must be care in the sequential computation. Due to the applications of the Kirkwood superposition approximation and the sequential calculation process, errors in $g_1$ originated from the early calculation process are apt to be accumulated in the reset of $g_1$. Hence, the values of $g_1$ far from the solid plate tend to deviate from 1. The deviation is very large, especially when the solid plate is strongly solvophobic or solvophilic. In that case, it is recommended to change the value of $g_C$ until the values of $g_1$ far from the solid plate become 1. (You do not have to prepare a new parameter for $g_C$ in the correction, because the values of $g_1$ far from the solid plate adjust the $g_C$.) After the correction, the goal of this short letter, the solvation structure ($\rho_0 g_1$) is obtained. This kind of retrieval can be performed by using $P$, $A$, or $PA$ instead of changing $g_C$. Also in this case, the parameter is not changed artificially, but mathematically. The values of $g_1$ far from the solid plate (= 1) adjust the parameter.

After calculation of $g_1$, one can calculate the structure of the confined liquid between the two solid plates by using the Kirkwood superposition approximation.

Since current experimental techniques (e.g., X-ray and neutron diffractions) cannot measure the confined liquid's structure, our technique is considered to be helpful for study of the confined liquid. It is considered that an elucidation of the structure of the confined liquid is beneficial to development of nanotribology.

## 3. Conclusions

We have proposed an improved method for transforming a force curve measured by SFA into the solvation structure on the surface of the SFA probe. The cylindrical solids are immersed in the solvent, where two-body potential between the solvent spheres is arbitrary. The improved point is that two-body potential between the SFA probe and the solvent sphere is modeled as attractive soft potential and the rigid wall. Thus, the transform method becomes more realistic compared with the previous work [3]. The transform method has been derived based on the statistical mechanics of the simple liquids, where we have taken advantage of the Kirkwood super position approximation.

In the near future, we will apply the transform method to the real SFA measurement in order to obtain the solvation structure on a surface *experimentally*. Although the transform method is constructed in the simple liquid, it is considered that the transform method can also be used for calculation of the density distribution of colloid particles on a surface. Moreover, it can also be used for measurement of the wettability of a surface, because the wettability is related to the solvation structure. The transform method is now on a stage of our computational verification. The results of the computational verification will be shown in the near future.

**ACKNOWLEDGEMENTS**
We appreciate discussions with M. Kisnoshita, H. Onishi, T. Sakka, N. Nishi, and O. Takahashi.